# Characterization of half-metallic $L2_1$-phase $Co_2FeSi$ full-Heusler alloy films formed by rapid thermal annealing

Yota Takamura [a]

*Department of Electronics and Applied Physics, Tokyo Institute of Technology, 4259 Nagatsuta, Midori-ku, Yokohama 226-8503, Japan*

Ryosho Nakane

*Department of Electronic Engineering, The University of Tokyo, 7-3-1 Hongo, Bunkyo-ku, Tokyo 113-8656, Japan*

Hiro Munekata and Satoshi Sugahara [b]

*Imaging Science and Engineering Laboratory, Tokyo Institute of Technology, 4259 Nagatsuta, Midori-ku, Yokohama 226-8503, Japan, and Department of Electronics and Applied Physics, Tokyo Institute of Technology, 4259 Nagatsuta, Midori-ku, Yokohama 226-8503, Japan*

---

[a] Electronic mail: yota@isl.titech.ac.jp

[b] Electronic mail: sugahara@isl.titech.ac.jp

ABSTRACT

The authors developed a preparation technique of $Co_2FeSi$ full-Heusler alloy films with the $L2_1$-ordered structure on silicon-on-insulator (SOI) substrates, employing rapid thermal annealing (RTA). The $Co_2FeSi$ full-Heusler alloy films were successfully formed by RTA-induced silicidation reaction between an ultrathin SOI (001) layer and Fe/Co layers deposited on it. The highly (110)-oriented $L2_1$-phase polycrystalline full-Heusler alloy films were obtained at the RTA temperature of 700 ºC. Crystallographic and magnetic properties of the RTA-formed full-Heusler alloy films were qualitatively the same as those of bulk full-Heusler alloy. This technique is compatible with metal source/drain formation process in advanced CMOS technology and would be applicable to the fabrication of the half-metallic source/drain of MOSFET type of spin transistors.

BODY

In recent years, spin transistors[1-7] attract considerable attention, since they have interesting transistor behavior with a new degree of freedom in controlling output currents based on spin-related phenomena. This feature makes it possible that spin transistors have useful nonvolatile and reconfigurable functionalities to improve circuit performance beyond present transistors. However, high current drive capability and a large on/off current ratio are still important requirements even for spin transistors, as well as present ordinary transistors[7]. In addition, excellent scalability and integration ability are also indispensable to establish very large scale integration. From these points of view, one of the promising spin transistors is recently proposed spin metal-oxide-semiconductor field-effect transistors (spin MOSFETs)[4-7]. The basic structure of spin MOSFETs consists of a MOS capacitor and ferromagnet contacts for the source and drain, and the device performance depends on the material of the ferromagnetic source/drain. Half-metallic ferromagnets[8] (HMFs) are the most effective source/drain material to realize the spin-dependent output characteristics of spin MOSFETs[4,5], since the extremely high spin polarization and unique spin-dependent band structure of HMFs possibly achieve spin injection into a semiconductor channel and spin detection from it with high efficiency.

Recently emerging silicide-based metal source/drain MOSFET[9-11] technology would be applicable to the fabrication of spin MOSFETs, owing to the similarity in the device structure between metal source/drain MOSFETs and spin MOSFETs. For metal source/drain MOSFETs, transition metal (or rare earth metal) silicides are directly used as their source and drain. These silicides are formed by thermally activated intermixing reaction, the so-called silicidation, between a Si substrate and transition metal (or rare earth metal) layer deposited on the Si substrate. In general, the thermal process for silicidation is induced by rapid thermal annealing (RTA). $L2_1$-phase half-metallic full-Heusler alloys containing Si ($Co_2FeSi$[12,13], $Co_2MnSi$[14,15,16], etc.) are one of attractive candidates for spin MOSFETs with the HMF source/drain, since these materials are considered to be a kind of silicides and thus they have a possibility to be formed by the RTA-induced silicidation process. In this paper, we developed a RTA-induced silicidation technique to form $L2_1$-phase $Co_2FeSi$ full-Heusler alloy films, utilizing a silicon-on-insulator (SOI) substrate. Magnetism in the RTA-formed $Co_2FeSi$ full-Heusler alloy films is also discussed.

The problem of the RTA technique for ferromagnetic silicides including Si-containing full-Heusler alloys is the formation of non-magnetic silicides with low magnetic element content. This is caused by the deep diffusion of magnetic elements into a Si substrate. In other words, when the much amount of Si atoms is supplied from the Si substrate, the most thermodynamically favorable nonmagnetic silicides are preferentially formed. Figure 1 shows the proposed preparation method of full-Heusler alloys employing a SOI substrate. The diffusion of magnetic atoms is blocked by the buried oxide (BOX) layer of the SOI substrate, and thus the composition of the RTA-formed film layer can be easily controlled by the thickness of the magnetic metal and SOI layers. When the amount of constituent elements of the resulting film layer is limited (or adjusted) to the chemical composition of full-Heusler alloys, homogeneous full-Heusler alloys would be formed, since full-Heusler alloys are also one of thermodynamically stable phases as long as keeping their (near) stoichiometric composition. Recently, ferromagnetic $Fe_3Si$ that has $L2_1$-related $DO_3$ structure were successfully formed by this technique[17].

All the experiments were carried out using commercially available wafer-bonding SOI substrates that consist of a thin SOI layer, a BOX layer and a Si substrate (hereafter, this structure is referred to as SOI/BOX/Si). A SOI substrate sample was cleaned by chemical oxidation using $H_2SO_4$ and $H_2O$ mixture and successive etching of the resulting surface oxide by buffered HF solution in order to eliminate carbon- and metal-related contaminations. The sample was transferred into an electron-beam evaporation chamber using an oil-free load-lock system, and then Co and Fe films were deposited on the SOI layer surface with a rate of 2.8nm/min in an ultrahigh vacuum. Subsequently, silicidation was performed by RTA in $N_2$ atmosphere. The thicknesses of Co and Fe films were measured by an uncalibrated quartz oscillator thickness monitor. Assuming the hexagonal close-packed (hcp) structure of Co and the body centered cubic (bcc) structure of Fe, the chemical composition of the sample was deduced from the film thicknesses of the Co, Fe, and SOI layers. The sample structure and a central experimental condition were as follows: the SOI thickness $d_{SOI}$ of 40 nm, the Co thickness $d_{Co}$ of 45 nm, the Fe thickness $d_{Fe}$ of 24 nm, the annealing temperature $T_A$ of 700 ºC, and the annealing time $t_A$ of 4 min. These parameters were used throughout the following experiments, unless otherwise noted. Hereafter, $Co_2FeSi$ will be expressed as CFS even for off-stoichiometric composition.

Figure 2(a) shows depth profiles of Co, Fe, Si and O in a RTA-treated Fe/Co/SOI/BOX/Si sample. The depth profiles were measured by secondary ion mass spectroscopy (SIMS) with the MCs+ technique[18], and the SIMS intensity of each element was calibrated by Rutherford backscattering and particle induced x-ray emission measurements. As expected, the diffusion of Co and Fe was completely blocked by the BOX layer, and the Co, Fe and Si concentrations in the RTA-formed film exhibited plateau profiles. The concentrations of Co, Fe and Si were 48 %, 30 %, and 22 %, respectively, which was slightly off-stoichiometric composition. The deviation from the stoichiometric composition would be caused by the uncalibrated quartz oscillator thickness monitor. On the other hand, when the silicidation was performed using an ordinary Si substrate instead of a SOI substrate, Fe and Co were much deeply diffused into the Si substrates as shown in Fig. 2(b), leading to the significant decrease and less controllability of the Fe and Co concentrations.

The crystallographic features of the RTA-formed CFS film were characterized by X-ray diffraction (XRD). After the Fe/Co/SOI/BOX/Si sample was annealed at $T_A$ = 700 ºC, the RTA-formed film showed strong CFS (220) and (440) diffraction peaks and no other diffraction peaks of CFS were observed in the $2\theta$ range between 20 º and 120 º, indicating that the RTA-formed CFS thin film was highly (110)-oriented. Note that the lattice constant of the CFS film was 0.5636 nm, which was slightly smaller than that of the bulk value (0.5647 nm) [19]. The deviation of the lattice constant from the bulk value would be caused by the off-stoichiometric composition. When an ordinary Si substrate was used, the sample showed no CFS-related diffraction peak, as shown by the top curve in Fig.

3(a).  These results were consistent with the above-described SIMS observations. The thermally activated silicidation process in the CFS formation was also traced by XRD measurements.  The results are also shown in Fig. 3(a).  An as-deposited Fe/Co/SOI/BOX/Si sample showed no diffraction peak at around the CFS (220) diffraction ($2\theta \sim 45.6\,°$).  When the sample was annealed at $T_A = 400\,°C$, a diffraction peak newly appeared at $2\theta = 45.0\,°$, which was slightly smaller than the peak position of the CFS (220) diffraction. This diffraction peak can be assigned by $Co_9Fe_9Si_2$ (110) or $FeSi_2$ (421), indicating insufficient silicidation reaction.  The diffraction of CFS (220) started to be observed at $T_A = 600\,°C$, resulting in the bimodal shape of the diffraction pattern, i.e., the film contained at least two crystalline phases.  Above $T_A = 700\,°C$, the diffraction peak at $45.0\,°$ was vanished. Only the CFS (220) diffraction was clearly observed and no other obvious diffraction peaks were detected, as described above.  When the direction normal to the sample plane was inclined toward $45\,°$ and $35.3\,°$ from the x-ray-source–detector–sample plane, the CFS (200) and (111) diffractions indicating fully ordered $L2_1$ structures were clearly observed for the RTA-formed CFS film, as shown in Fig. 3(b).  It was concluded from above described SIMS and XRD measurements that highly (110)-oriented $L2_1$-ordered CFS films were successfully formed by RTA-induced silicidation, using a SOI substrate.

Figure 4 shows the saturation magnetization $M_S$ and coercivity $H_C$ of the RTA-formed CFS thin films as a function of $T_A$, where $M_S$ was normalized by the saturation magnetization $M_S^{Co,Fe}$ of their as-deposited sample.  The measurements were carried out using a superconductive quantum interference device (SQUID) magnetometer at 300 K.  All the data were taken with a magnetic field applied parallel to the in-plane direction of the SOI (001) substrate. The changes in $M_S$ and $H_C$ can be divided into three regions, as noted in the figure. In the region I (300 °C ≤ $T_A$ ≤ 400 °C), $M_S/M_S^{Co,Fe}$ maintained almost unity and $H_C$ little changed, indicating that there is no detectable silicidation reaction. In the region II (400 °C ≤ $T_A$ ≤ 650 °C), $M_S$ rapidly decreased and took a minimum value. This implies that the thermally activated silicidation process proceeded, and that the formation of nonmagnetic silicides, such as $FeSi_2$, reduced $M_S/M_S^{Co,Fe}$. In this region, $H_C$ took a maximum value at around $T_A$ = 500 °C, and then deceased with increasing $T_A$, which would also reflect the uniformity of the film. In the region III (650 °C ≤ $T_A$ ≤ 800 °C), $M_S$ increased to ~$M_S^{Co,Fe}$ and then was slightly reduced with increasing $T_A$. In this region, $H_C$ took a almost constant value of 89 Oe. In addition, the $M$-$H$ curves of the samples were a single-step square-like hysteresis loop with sharp magnetization switching. These results mean that when the L2$_1$-phase CFS film was formed, the magnetic properties were drastically improved. The magnetic moment per unit cell of the sample in the region III was 5.2 $\mu_B$ – 4.8 $\mu_B$, which is slightly smaller than the bulk value (6 $\mu_B$)[13]. This might be caused by the deviation of chemical composition from stoichiometric CFS, and by the incorporation of oxygen owing to a residual oxide layer on the SOI substrate before the deposition of Co and Fe.

    A solid curve in Fig. 5 shows the magnetic circular dichroism (MCD) spectrum of the RTA-formed CFS film ($T_A$ = 700 °C). MCD spectra for its as-deposited Fe/Co/SOI/BOX/Si sample (whose surface is Fe), an as-deposited Co/SOI/BOX/Si sample, and a RTA-treated Fe/Co/SiO$_2$/Si sample with $T_A$ = 700 °C are also shown in the figure as references. All the spectra were measured at

room temperature with a reflection configuration with a magnetic field $H$ of 1 T applied perpendicular to the sample plane. In general, a MCD spectrum sensitively reflects the band structure of a material, since MCD is given by the difference in optical reflectance (absorption) between right- and left-circular polarized lights. Therefore, all the reference samples showed different spectral features, as shown in Fig. 5. Note that the MCD spectrum of the as-deposited Fe/Co/SOI/BOX/Si sample was qualitatively the same as that of previously reported theoretical and experimental results of Fe thin films[20,21]. The spectral features of the RTA-formed CFS were clearly different from those of reference samples, indicating that the formation of the $L2_1$-phase full-Heusler alloy by the RTA-induced silicidation significantly modified its band structure. In addition, the MCD spectra of the RTA-formed CFS film measured with the several magnitudes of the applied magnetic field exhibited identical magnetic field dependence, i.e., when the intensity of these spectra were normalized, these spectra were completely overlapped with one another. This indicates that the RTA-formed CFS film was magnetically homogeneous, in other words, ferromagnetic precipitates (residual Co, Fe and other ferromagnetic silicides) can be excluded for the origin of the ferromagnetism in the CFS film.

In summary, $L2_1$-phase CFS full-Heusler alloy films were successfully formed by RTA-induced silicidation, utilizing a SOI substrate. This technique is fully compatible with metal source/drain process in advanced CMOS technology and thus is promising for the fabrication of MOSFET-based spin transistors.

The authors would like to thank Profs. M. Tanaka and S. Takagi, The University of Tokyo. This work was in part supported by Industrial Technology Research Grant Program from NEDO.

Figure captions

Fig. 1 Schematic illustration of Co$_2$FeSi (CFS) formation process using a silicon-on-insulator (SOI) substrate.

Fig. 2 SIMS depth profiles of Co, Fe, Si and O (a) in a RTA-treated Fe/Co/SOI/BOX/Si sample with $T_A$ = 700 °C, and (b) in a RTA-treated Fe/Co/Si sample with $T_A$ = 700 °C.

Fig. 3(a) XRD patterns of the RTA-trated Fe/Co/SOI/BOX/Si samples with various RTA-temperatures ($T_A$). The XRD pattern of a RTA-treated Fe/Co/Si sample with $T_A$ = 700 °C is also shown as a reference (topmost XRD pattern). (b) Superlattice diffraction of the RTA-formed CFS film with $T_A$ = 700 °C. $\varphi$ is the tilted angle of the sample plane.

Fig. 4 Magnetization $M_S$ /$M_S^{Co,Fe}$ (solid circles) and coercivity $H_C$ (open circles) as a function of RTA-temperature $T_A$ for the RTA-formed CFS samples.

Fig. 5 MCD spectra of the RTA-formed CFS film with $T_A$ = 700 ºC (solid line), its as-deposited Fe/Co/SOI/BOX/Si sample (broken line), an as-deposited Co/SOI/BOX/Si sample (broken-dotted line) and a RTA-treated Fe/Co/SiO$_2$/Si sample (dotted line) with $T_A$ = 700 °C, measured at room temperature with a magnetic field of 1 T applied perpendicular to the sample plane.

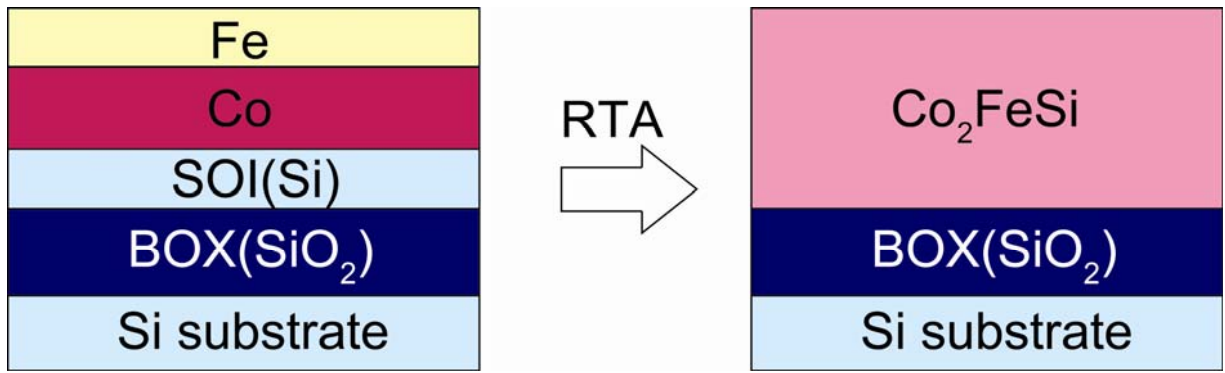

Figure 1 (Color online) Takamura *et al*.

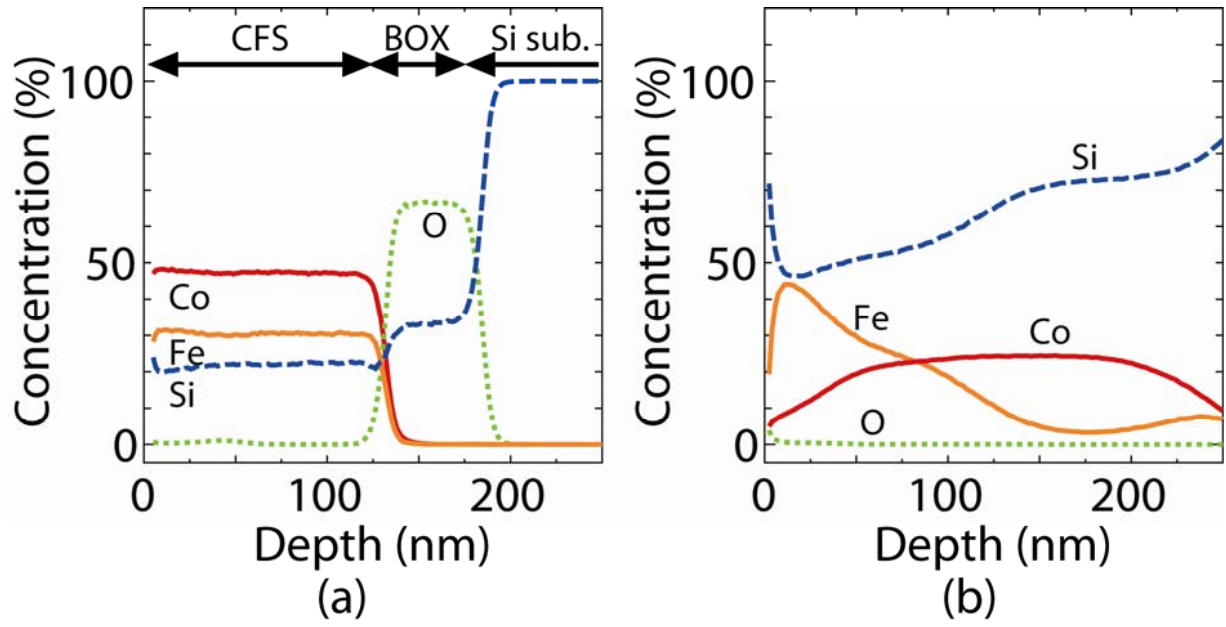

Figure 2 (Color online) Takamura *et al*.

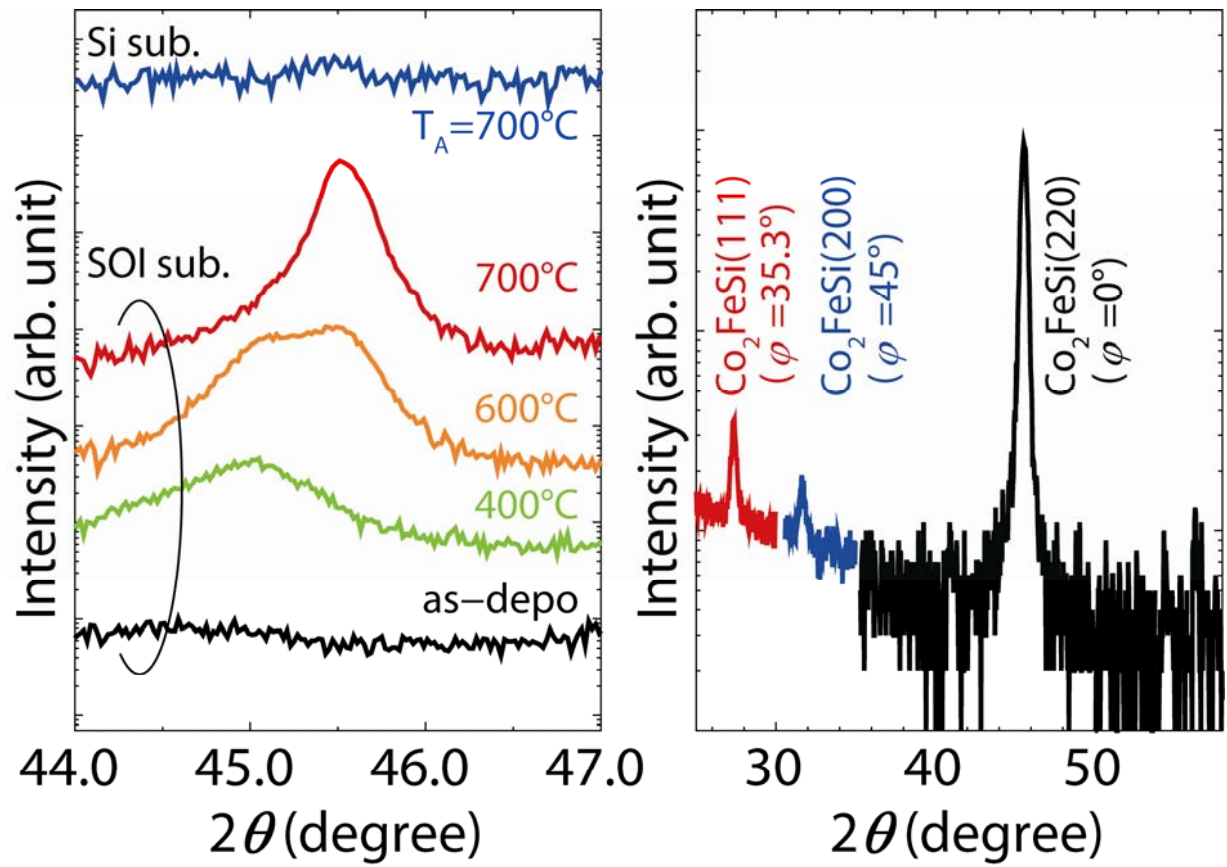

Figure 3 (Color online) Takamura *et al*.

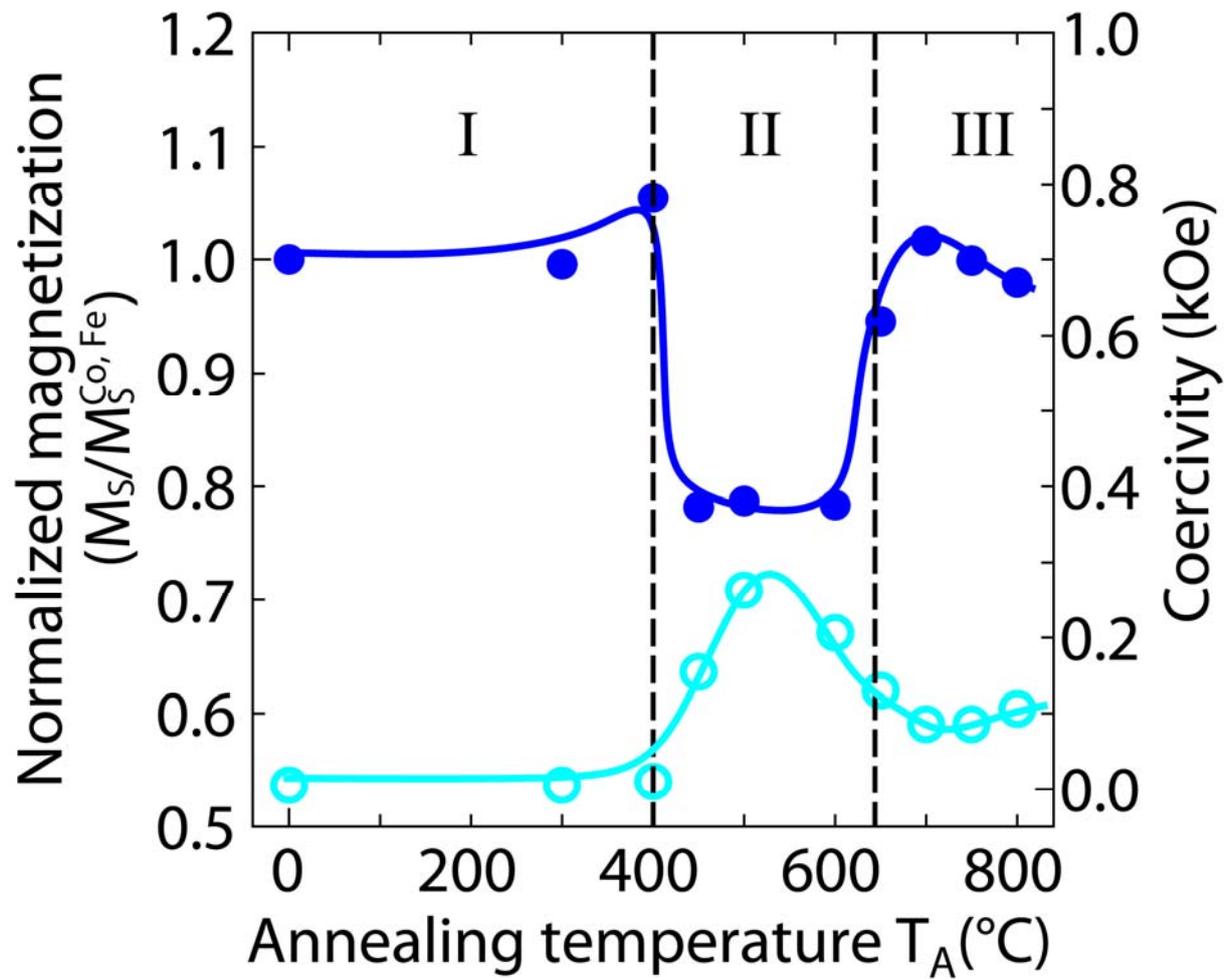

Figure 4 (Color online) Takamura *et al*.

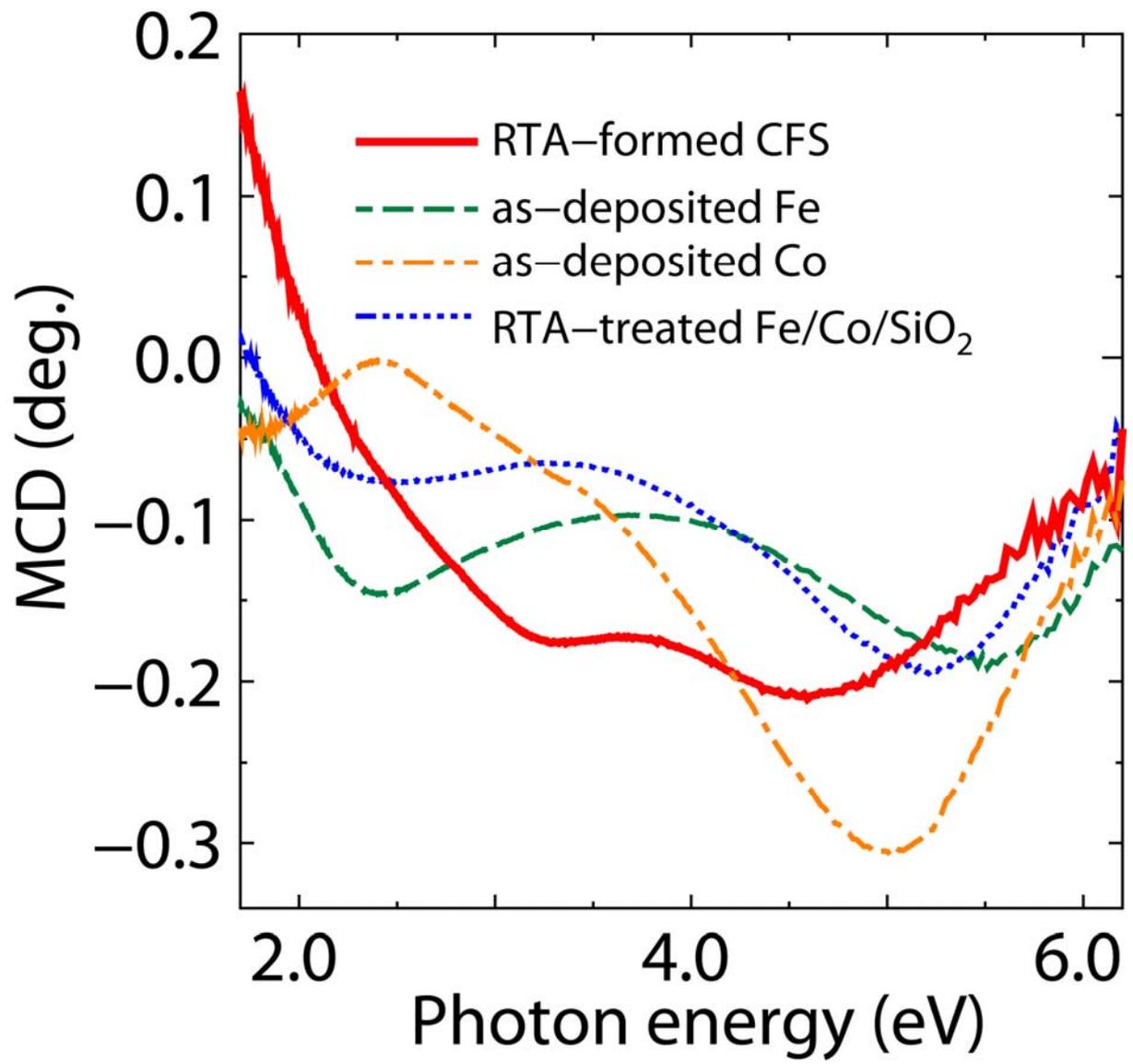

Figure 5 (Color online) Takamura *et al*.